\documentclass[ifip]{svmult}

\usepackage{makeidx}         
\usepackage{multicol}        
\usepackage[bottom]{footmisc}
\usepackage{cite}            

\makeindex             

\begin{document}

\title*{Unreduced Dynamic Complexity:\\
Towards the Unified Science of Intelligent Communication Networks
and Software}

\titlerunning{Unreduced Dynamic Complexity of Intelligent Communication Networks}

\author{Andrei P. Kirilyuk}

\authorrunning{A.P. Kirilyuk}

\institute{Solid State Theory Department, Institute of Metal
Physics\\36 Vernadsky Avenue, 03142 Kiev-142, Ukraine\\
\texttt{kiril@metfiz.freenet.kiev.ua}}
%
%

\maketitle

Operation of autonomic communication network with complicated
user-oriented functions should be described as unreduced many-body
interaction process. The latter gives rise to complex-dynamic
behaviour including fractally structured hierarchy of chaotically
changing realisations. We recall the main results of the universal
science of complexity based on the unreduced interaction problem
solution and its application to various real systems, from
nanobiosystems and quantum devices to intelligent networks and
emerging consciousness. We concentrate then on applications to
autonomic communication leading to fundamentally substantiated,
exact science of intelligent communication and software. It aims
at unification of the whole diversity of complex information
system behaviour, similar to the conventional, ``Newtonian"
science order for sequential, regular models of system dynamics.
Basic principles and first applications of the unified science of
complex-dynamic communication networks and software are outlined
to demonstrate its advantages and emerging practical perspectives.

\section{Introduction}\label{Intro}
Although any information processing can be described as interaction
between participating components, communication and software tools
used today tend to limit this interaction to \emph{unchanged},
preprogrammed system configuration, which reduces underlying
interaction processes to a very special and rather trivial type of a
sequential and regular ``Turing machine". It does not involve any
genuine novelty emergence (with the exception of thoroughly avoided
failures), so that the occurring ``normal" events do not change
anything in system configuration and are supposed to produce only
\emph{expected} modification in particular register content, etc. In
terms of fundamental science, one deals here with a rare limiting
case of exactly solvable, \emph{integrable} and certainly
\emph{computable} interaction problem and dynamics. The latter
property, associated with total \emph{regularity},
\emph{predictability}, and \emph{decidability}, forms even a major
\emph{purpose} and invariable operation principle of all traditional
information processing and communication systems (it can be
generalised to all man-made tools and engineering approaches). One
can say also that usual ICT systems do not possess any genuine
\emph{freedom}, or unreduced, \emph{decision-taking} autonomy, which
is totally displaced towards the human user side. One should not
confuse that truly autonomic, \emph{independent} kind of action with
a high degree of man-made \emph{automatisation} that can only
(usefully) \emph{imitate} a decision making process, while
preserving the basic predictability and computability of such
``intelligent" machine behaviour.

In the meanwhile, unreduced interaction processes involving
essential and \emph{noncomputable} system configuration change
conquer practically important spaces at a catastrophically growing
rate. On one hand, this rapid advance of unreduced interaction
complexity occurs inevitably in communication networks and related
software themselves as a result of strong increase of user numbers,
interests, and functional demands (i.e. desired ``quality of
service"). On the other hand, it is due to simultaneous, and
related, growth of popularity of \emph{complex-dynamical
applications} of ICT tools, which can be truly efficient only if
those tools possess corresponding levels of \emph{unreduced,
interaction-driven complexity} and autonomy. It can appear either as
unacceptably growing frequency of complicated system
\emph{failures}, or else as the transition to a fundamentally
different operation mode that uses inevitable uncertainties of
complex interaction dynamics to strongly \emph{increase system
performance quality}.

The second aspect has its deep, objective origin in today's specific
moment of development, where one can \emph{empirically} modify,
\emph{for the first time} in history, the \emph{whole} scale of
natural system complexity at \emph{all} its levels, from ultimately
small, quantum systems (high-energy physics) to the most
complicated, biological, ecological, and conscious systems
(genetics, industrial transformation, social and psychological
changes). However, the genuine \emph{understanding} of complex
dynamics we \emph{strongly modify} is persistently \emph{missing}
from modern science, which gives rise to multiple and real
\emph{dangers} (c.f. \cite{Rees}) and determines the \emph{urgency}
of \emph{complexity transition} in major ICT \emph{instruments of
progress}. In fact, any progress becomes blocked when
zero-complexity, regular computing instruments encounter the
unreduced, \emph{noncomputable} complexity of real systems.

Those reasons determine increased attention to autonomic, complex,
``bio-inspired" operation of communication networks and software
tools that appears within various overlapping research initiatives,
such as ``pervasive computing", ``ambient intelligence", ``autonomic
communication networks", ``knowledge-based networks", ``context
awareness", ``semantic grid/web", ``complex software", etc. (see
e.g. \cite{WAC,FP6,Bull,DiMarzo,SemWeb,ManyOne,KocarevVattay} for
overview and further references). In fact, such kind of development
has become a major, rapidly growing direction of ICT research.
However, as big volumes of ``promising" results and ``futuristic"
expectations accumulate, the need for a \emph{unified} and
\emph{rigorous} framework, similar to exact science for non-complex,
regular models, becomes evident.

Although creation of ``intelligent" ICT structures has
\emph{started} from applied, \emph{engineering} studies
(\emph{contrary} to traditional science applications), the emerging
\emph{qualitative} novelties call for a \emph{new kind} of
\emph{fundamental knowledge} as a basis for those practically
oriented efforts, without which there is a \emph{serious risk} to
``miss the point" in unlimited diversity of unreliable guesses. The
created \emph{proto-science} state of autonomic communication and
complex software should now be extended to the \emph{genuine
scientific knowledge} of a new kind. This \emph{exact science}
should provide \emph{rigorously specified and unified} description
of \emph{real}, not ``model", complex systems with \emph{unreduced}
interaction of ``independent" components, leading to
\emph{explicitly emerging}, ``unexpected" structures and properties.

One can call such unreduced interaction process \emph{generalised
autonomic network}, even though it is realised eventually in
\emph{any} kind of unreduced interaction process, including
\emph{complex software} systems. We deal here with an important
quality of new knowledge, absent in the model-based and
intrinsically split usual science: once the \emph{truly consistent
understanding} of real complex interaction is established, it should
\emph{automatically} be applicable to any other, maybe externally
quite different case of equally complex many-body
interaction.\footnote{This difference from the usual, ``Newtonian"
science (including its versions of ``new physics" and ``science of
complexity") stems from the fact that it does \emph{not} propose any
\emph{solution} to the \emph{unreduced}, realistic interaction
problem and does not really aim at obtaining the related
\emph{complete explanation} of the analysed phenomena, replacing it
with a mathematically ``exact" (analytically ``closed"), but
extremely simplified, unrealistic, and often guessed ``models" and
``postulates", which are then mechanically adjusted to separate,
subjectively chosen observation results.} Thus, truly autonomic
communication networks may show basically the same patterns of
behaviour and evolution as full-scale nanobiosystems, biological
networks in living organisms and ecosystems, or intelligent neural
networks. It is a fortunate and non-occasional circumstance, since
those complex-dynamic ICT tools will be properly suited to their
material basis and application objects, complex systems from the
``real world", so that finally there is \emph{no separation and
basic difference} between complex behaviour of ``natural",
real-world and ``artificial", ICT systems: \emph{everything} tends
to be irreducibly complex, interconnected, \emph{and} man-made,
whereupon the necessity of a unified and realistic understanding of
emerging new, \emph{explicitly complex} world becomes yet more
evident.
\index{autonomic network}
\index{complex software}
\index{autonomic communication}

Note again the essential difference with the prevailing usual
science approach, where the unreduced complexity of real structures
is \emph{artificially} reduced, i.e. \emph{completely destroyed}, so
that ``computer simulation" of such basic structures as elementary
particles or biomolecules takes the highest existing
``supercomputer" (or grid) powers and still cannot provide a really
useful, unambiguous result. In view of such situation one can have
justified doubts in the ability of usual, scholar science to
describe truly complex systems without losing the underlying rigour
(objectivity) of the scientific method itself (see e.g.
\cite{Rees,Perplexity,EndScience,Ellis:1,Ellis:2}).

In this report we describe a theory of arbitrary many-body
interaction leading to a universally applicable concept of dynamic
complexity
\cite{Kir:USciCom,Kir:SelfOrg,Kir:SymCom,Kir:Fract:1,Kir:RealQuMach,Kir:Fract:2,Kir:Nano,Kir:Conscious,Kir:CommNet,Kir:SustTrans,Kir:QFM,Kir:Cosmo,Kir:QuChaos,Kir:Channel}
(Sect.~\ref{ComDyn}). The resulting \emph{universal science of
complexity} provides a working prototype of the new kind of science
claimed by modern development of information and communication
systems \cite{Kir:CommNet}. We then specify its application to
emerging complex software and autonomic communication systems
(Sect.~\ref{SciComNet}). We finally summarise the key features of
the ``new mathematics of complexity" as a fundamental and rigorous
basis for the new science of intelligent information systems
(Sect.~\ref{NewMath}) and draw major development perspectives
(Sect.~\ref{Conclusion}).

\section{Complex Dynamics of Unreduced Interaction Process}\label{ComDyn}
\subsection{Multivalued Dynamics and Unified Complexity Concept}\label{UniCom}
We begin with a general equation for arbitrary system dynamics (or
many-body problem), called here \emph{existence equation} and simply
fixing the fact of interaction between the system components (it
generalises various model equations):
\begin{equation}\label{eq:1}
\left\{ {\sum\limits_{k = 0}^N {\left[ {h_k \left( {q_k } \right) +
\sum\limits_{l > k}^N {V_{kl} \left( {q_k ,q_l } \right)} } \right]}
} \right\}\Psi \left( Q \right) = E\Psi \left( Q \right)\ ,
\end{equation}
where $h_k \left( {q_k } \right)$ is the ``generalised Hamiltonian"
of the $k$-th system component, $q_k$ is the degree(s) of freedom of
the $k$-th component, $V_{kl} \left( {q_k ,q_l } \right)$ is the
(arbitrary) interaction potential between the $k$-th and $l$-th
components, $\Psi \left( Q \right)$ is the system state-function, $Q
\equiv \left\{ {q_0 ,q_1 ,...,q_N } \right\}$, $E$ is the
generalised Hamiltonian eigenvalue, and summations are performed
over all ($N$) system components. The generalised Hamiltonian,
eigenvalues, and interaction potential represent a suitable measure
of dynamic complexity defined below and encompassing practically all
``observable" quantities (action, energy/mass, momentum,
information, entropy, etc.). Therefore (\ref{eq:1}) can express
unreduced interaction configuration of arbitrary
communication/software system. If interaction potential (system
configuration) depends explicitly on time, one should use a
time-dependent form of (\ref{eq:1}), where eigenvalue $E$ is
replaced with the partial time derivative operator.

It is convenient to separate one of the degrees of freedom, e.g.
$q_0 \equiv \xi $, representing a naturally selected, usually
``system-wide" entity, such as component coordinates or ``connecting
agent" (here now $Q \equiv \left\{ {q_1 ,...,q_N } \right\}$ and
$k,l \ge 1$):
\begin{equation}\label{eq:2}
\left\{ {h_0 ( \xi ) + \sum\limits_{k = 1}^N {\left[ {h_k ( {q_k } )
+ V_{0k} ( {\xi ,q_k } )} + \sum\limits_{l
> k}^N {V_{kl} ( {q_k,q_l } )} \right]} } \right\}\Psi
( {\xi,Q} ) = E\Psi ( {\xi,Q} ),
\end{equation}

We express the problem in terms of known free-component solutions
for the ``functional", internal degrees of freedom of system
elements ($k \ge 1$):
\begin{equation}\label{eq:3}
h_k \left( {q_k } \right)\varphi _{kn_k } \left( {q_k } \right) =
\varepsilon _{n_k } \varphi _{kn_k } \left( {q_k } \right)\ ,
\end{equation}
\begin{equation}\label{eq:4}
\Psi \left( {\xi ,Q} \right) = \sum\limits_n {\psi _n \left( \xi
\right)} \varphi _{1n_1 } \left( {q_1 } \right)\varphi _{2n_2 }
\left( {q_2 } \right)...\varphi _{Nn_N } \left( {q_N } \right)
\equiv \sum\limits_n {\psi _n \left( \xi  \right)} \Phi _n \left( Q
\right),
\end{equation}
where $\left\{ {\varepsilon _{n_k } } \right\}$ are the eigenvalues
and $\left\{ {\varphi _{kn_k } \left( {q_k } \right)} \right\}$
eigenfunctions of the $k$-th component Hamiltonian $h_k \left( {q_k
} \right)$, forming the complete set of orthonormal functions, $n
\equiv \left\{ {n_1,...,n_N } \right\}$ runs through all possible
eigenstate combinations, and $\Phi _n \left( Q \right) \equiv
\varphi _{1n_1 } \left( {q_1 } \right)\varphi _{2n_2 } \left( {q_2 }
\right)...\varphi _{Nn_N } \left( {q_N } \right)$ by definition. The
system of equations for $\left\{ {\psi _n \left( \xi  \right)}
\right\}$ is obtained then in a standard way, using the
eigen-solution orthonormality (e.g. by multiplication by $\Phi _n^
* \left( Q \right)$ and integration over $Q$):
\begin{equation}\label{eq:5}
\begin{array}{lll}
\left[ {h_0 ( \xi ) + V_{00} ( \xi )} \right]\psi _0 ( \xi
)&+&\sum\limits_n {V_{0n} ( \xi )} \psi _n ( \xi ) = \eta \psi _0 (
\xi )\\
\left[ {h_0 ( \xi ) + V_{nn} ( \xi )} \right]\psi _n ( \xi
)&+&\sum\limits_{n' \ne n} {V_{nn'} ( \xi )} \psi _{n'} ( \xi ) =
\eta _n \psi _n ( \xi ) - V_{n0} ( \xi )\psi _0 ( \xi ),
\end{array}
\end{equation}
where $n,n' \ne 0$ (also below), $\eta \equiv \eta _0 = E -
\varepsilon _0$, $\eta _n = E - \varepsilon _n$, $\varepsilon _n =
\sum\limits_k {\varepsilon _{n_k } }$,
\begin{equation}\label{eq:6}
V_{nn'} \left( \xi  \right) = \sum\limits_k {\left[ {V_{k0}^{nn'}
\left( \xi  \right) + \sum\limits_{l > k} {V_{kl}^{nn'} } }
\right]}\ ,
\end{equation}
\begin{equation}\label{eq:7}
V_{k0}^{nn'} \left( \xi  \right) = \int\limits_{\Omega _Q } {dQ\Phi
_n^ *  \left( Q \right)} V_{k0} \left( {q_k ,\xi } \right)\Phi _{n'}
\left( Q \right)\ ,
\end{equation}
\begin{equation}\label{eq:8}
V_{kl}^{nn'} \left( \xi \right) = \int\limits_{\Omega _Q } {dQ\Phi
_n^ *  \left( Q \right)} V_{kl} \left( {q_k ,q_l } \right)\Phi _{n'}
\left( Q \right)\ ,
\end{equation}
and we have separated the equation for $\psi _0 ( \xi )$ describing
the generalised ``ground state" of system elements, i. e. the state
with minimum complexity. The obtained system of equations expresses
the same problem as the starting equation (\ref{eq:2}) but now in
terms of ``natural" variables, and therefore it results from various
starting models, including time-dependent and formally ``nonlinear"
ones.

We can try to solve the ``nonintegrable" system of equations
(\ref{eq:5}) with the help of generalised effective, or optical,
potential method \cite{Ded}, where one expresses $\psi _n \left( \xi
\right)$ through $\psi _0 \left( \xi \right)$ from equations for
$\psi _n \left( \xi  \right)$ using the standard Green function
technique and then inserts the result into the equation for $\psi _0
\left( \xi  \right)$, obtaining thus the \emph{effective existence
equation} that contains \emph{explicitly} only ``integrable" degrees
of freedom ($\xi$)
\cite{Kir:USciCom,Kir:SelfOrg,Kir:SymCom,Kir:Fract:1,Kir:RealQuMach,Kir:QuChaos,Kir:Channel}:
\begin{equation}\label{eq:9}
h_0 \left( \xi  \right)\psi _0 \left( \xi  \right) + V_{{\rm{eff}}}
\left( {\xi ;\eta } \right)\psi _0 \left( \xi \right) = \eta \psi _0
\left( \xi  \right)\ ,
\end{equation}
where the operator of \emph{effective potential (EP)},
$V_{{\rm{eff}}} \left( {\xi ;\eta } \right)$, is given by
\begin{equation}\label{eq:10}
V_{{\rm{eff}}} \left( {\xi ;\eta } \right) = V_{00} \left( \xi
\right) + \hat V\left( {\xi ;\eta } \right),\ \  \hat V\left( {\xi
;\eta } \right)\psi _0 \left( \xi  \right) = \int\limits_{\Omega
_\xi  } {d\xi 'V\left( {\xi ,\xi ';\eta } \right)} \psi _0 \left(
{\xi '} \right),
\end{equation}
\begin{equation}\label{eq:11}
V\left( {\xi ,\xi ';\eta } \right) = \sum\limits_{n,i}
{\frac{{V_{0n} \left( \xi  \right)\psi _{ni}^0 \left( \xi
\right)V_{n0} \left( {\xi '} \right)\psi _{ni}^{0*} \left( {\xi '}
\right)}}{{\eta  - \eta _{ni}^0  - \varepsilon _{n0} }}}\ ,\ \ \
\varepsilon _{n0}  \equiv \varepsilon _n  - \varepsilon _0\ ,
\end{equation}
and $\left\{ {\psi _{ni}^0 \left( \xi  \right)} \right\}$, $\left\{
{\eta _{ni}^0 } \right\}$ are complete sets of eigenfunctions and
eigenvalues of a \emph{truncated} system of equations:
\begin{equation}\label{eq:12}
\left[ {h_0 \left( \xi  \right) + V_{nn} \left( \xi  \right)}
\right]\psi _n \left( \xi  \right) + \sum\limits_{n' \ne n} {V_{nn'}
\left( \xi  \right)} \psi _{n'} \left( \xi  \right) = \eta _n \psi
_n \left( \xi  \right)\ .
\end{equation}

Since the unreduced EP (\ref{eq:10})--(\ref{eq:11}) depends
essentially on the eigen-solutions to be found, the problem remains
``nonintegrable" and formally equivalent to the initial formulation
(\ref{eq:1}),(\ref{eq:2}),(\ref{eq:5}). However, it is the effective
version of a problem that reveals the nontrivial properties of its
unreduced solution. The most important property of the unreduced
interaction result (\ref{eq:9})--(\ref{eq:12}) is its \emph{dynamic
multivaluedness} meaning that one has a \emph{redundant} number of
different but individually complete, and therefore \emph{mutually
incompatible}, problem solutions describing \emph{equally real}
system configurations. We therefore call each of them
\emph{realisation} of the system and problem. Plurality of system
realisations follows from the \emph{dynamically nonlinear} EP
dependence on the solutions to be found, reflecting the evident
plurality of interacting eigen-mode combinations
\cite{Kir:USciCom,Kir:SelfOrg,Kir:SymCom,Kir:Fract:1,Kir:RealQuMach,
Kir:Fract:2,Kir:Nano,Kir:Conscious,Kir:CommNet,Kir:SustTrans,Kir:QFM,
Kir:Cosmo,Kir:QuChaos,Kir:Channel}.\index{dynamic multivaluedness}

It is important that dynamic multivaluedness emerges only in the
unreduced problem formulation, whereas the standard theory,
including usual EP method applications (see e.g. \cite{Ded}) and the
scholar ``science of complexity" (theory of chaos,
self-organisation, etc.), resorts invariably to one or another
version of \emph{perturbation theory}, whose ``mean-field"
approximation, providing an ``exact", closed-form solution, totally
kills dynamic redundance by eliminating the nonlinear dynamical
links in (\ref{eq:9})--(\ref{eq:11}) and retaining \emph{only one},
``averaged" solution, usually expressing but small deviations from
\emph{imposed} system configuration:
\begin{equation}\label{eq:13}
\left[ {h_0 \left( \xi  \right) + V_{nn} \left( \xi  \right) +
\tilde V_n \left( \xi  \right)} \right]\psi _n \left( \xi  \right) =
\eta _n \psi _n \left( \xi  \right)\ ,
\end{equation}
where $ \left| {V_0 \left( \xi  \right)} \right|  < \left| {\tilde
V_n \left( \xi  \right)} \right| < \left| {\sum\limits_{n'} {V_{nn'}
} \left( \xi  \right)} \right| $. General problem solution is then
obtained as an essentially linear \emph{superposition} of
eigen-solutions of (\ref{eq:13}) similar to (\ref{eq:4}). This
\emph{dynamically single-valued}, or \emph{unitary}, problem
reduction forms the basis of the whole canonical science paradigm.

The \emph{unreduced}, truly complete \emph{general solution} to a
problem emerges as a \emph{dynamically probabilistic} sum of
\emph{redundant} system \emph{realisations}, each of them being
roughly equivalent to the whole ``general solution" of usual theory:
\begin{equation}\label{eq:14}
\rho \left( {\xi ,Q} \right) = \sum\limits_{r = 1}^{N_\Re  } {^{^
\oplus}  \rho _r \left( {\xi ,Q} \right)}\ ,
\end{equation}
where the observed (generalised) density, $\rho \left( {\xi ,Q}
\right)$, is obtained as the state-function squared modulus, $\rho
\left( {\xi ,Q} \right) = \left| {\Psi \left( {\xi ,Q} \right)}
\right|^2 $ (for ``wave-like" complexity levels), or as the
state-function itself, $\rho \left( {\xi ,Q} \right) = \Psi \left(
{\xi ,Q} \right)$ (for ``particle-like" structures), index $r$
enumerates system realisations, $N_\Re$ is realisation number (its
maximum value is equal to the number of system components, $N_\Re =
N$), and the sign $\oplus$ designates the special, dynamically
probabilistic meaning of the sum. The latter implies that
incompatible system realisations are forced, by the \emph{same}
driving interaction, to \emph{permanently replace each other} in a
\emph{causally (dynamically) random order} thus consistently
defined. The $r$-th realisation state-function, $\Psi _r \left( {\xi
,Q} \right)$, in the unreduced general solution (\ref{eq:14}) is
obtained as
\[
\Psi _r \left( {\xi ,Q} \right) = \sum\limits_i {c_i^r } \left[
{\Phi _0 \left( Q \right)\psi _{0i}^r \left( \xi  \right)} \right. +
\]
\begin{equation}\label{eq:15}
+ \sum\limits_{n, i'} {\left. {\frac{{\Phi _n \left( Q \right)\psi
_{ni'}^0 \left( \xi  \right)\int\limits_{\Omega _\xi } {d\xi '\psi
_{ni'}^{0*} \left( {\xi '} \right)V_{n0} \left( {\xi '} \right)\psi
_{0i}^r \left( {\xi '} \right)} }}{{\eta _i^r  - \eta _{ni'}^0  -
\varepsilon _{n0} }}} \right] }\ ,
\end{equation}
where $\{ \psi _{0i}^r \left( \xi  \right),\eta _i^r \}$ are $r$-th
realisation eigen-solutions of the unreduced EP equation
(\ref{eq:9}) and the coefficients $c_i^r$ should be found from the
state-function matching conditions at the boundary where interaction
effectively vanishes. The corresponding $r$-th realisation EP takes
the form (derived from (\ref{eq:10})--(\ref{eq:11})):
\[
V_{{\rm{eff}}} \left( {\xi ;\eta _i^r } \right)\psi _{0i}^r \left(
\xi  \right) = V_{00} \left( \xi  \right)\psi _{0i}^r \left( \xi
\right) +
\]
\begin{equation}\label{eq:16}
+ \sum\limits_{n, i'} {\frac{{V_{0n} \left( \xi  \right)\psi
_{ni'}^0 \left( \xi  \right)\int\limits_{\Omega _\xi  } {d\xi '\psi
_{ni'}^{0*} \left( {\xi '} \right)V_{n0} \left( {\xi '} \right)\psi
_{0i}^r \left( {\xi '} \right)} }}{{\eta _i^r  - \eta _{ni'}^0  -
\varepsilon _{n0} }}}\ .
\end{equation}
Equations (\ref{eq:14})--(\ref{eq:16}) reveal, in particular,
\emph{dynamic localisation} of a system in any its normal,
``regular" realisation around its characteristic eigenvalue and
configuration (due to the resonance denominator) and reverse
delocalisation during \emph{transition} between regular
realisations, occurring though a special, \emph{intermediate}
realisation of the \emph{wavefunction}
\cite{Kir:USciCom,Kir:RealQuMach,Kir:Conscious,Kir:QFM,Kir:75Wavefunc}
(see also below).

Direct comparison between the unreduced
(\ref{eq:9}),(\ref{eq:12}),(\ref{eq:14})--(\ref{eq:16}) and reduced
(\ref{eq:13}) problem solutions reveals the exact dynamic origin and
huge scale of difference between the real system complexity and its
model simplification in the unitary theory. In particular, the
unreduced solution (\ref{eq:14}) implies that any measured value is
\emph{intrinsically unstable} and \emph{will} unpredictably change
to another one, corresponding to another, \emph{randomly} chosen
realisation. Such kind of behaviour is readily observed in nature
and actually explains the living organism behaviour
\cite{Kir:USciCom,Kir:Fract:1,Kir:RealQuMach,Kir:Fract:2}, but is
thoroughly avoided in the unitary approach and technological systems
(including ICT systems), where it is correctly associated with
linear ``noncomputability" and technical failure (we shall consider
below that \emph{limiting} regime of complex dynamics). Therefore
the universal dynamic multivaluedness revealed by rigorous problem
solution forms the fundamental basis for the transition to
``bio-inspired" and ``intelligent" kind of operation in artificial,
technological and communication systems, where causal randomness can
be transformed from an obstacle to a qualitative advantage
(Sect.~\ref{SciComNet}).
\index{causal randomness}

The rigorously derived randomness of the generalised EP formalism
(\ref{eq:14})-(\ref{eq:16}) is accompanied by the \emph{dynamic
definition of probability}. As elementary realisations are
equivalent in their ``right to appear", the dynamically obtained,
\emph{a priori probability}, $\alpha _r$, of elementary realisation
emergence is given by
\begin{equation}\label{eq:17}
\alpha _r  = \frac{1}{{N_\Re  }}\ ,\ \ \ \sum\limits_r {\alpha _r }
= 1\ .
\end{equation}
However, a real observation may resolve only uneven groups of
elementary realisations. The dynamic probability of such general,
compound realisation is determined by the number, $N_r$, of
elementary realisations it contains:
\begin{equation}\label{eq:18}
\alpha _r \left( {N_r } \right) = \frac{{N_r }}{{N_\Re  }}\ \ \
\left( {N_r  = 1,...,N_\Re  ;\ \sum\limits_r {N_r }  = N_\Re  }
\right),\ \ \ \sum\limits_r {\alpha _r }  = 1\ .
\end{equation}
An expression for \emph{expectation value}, $\rho _{\exp } \left(
{\xi ,Q} \right)$, follows from
(\ref{eq:14}),(\ref{eq:17})--(\ref{eq:18}) for statistically long
observation periods:
\begin{equation}\label{eq:19}
\rho _{\exp } \left( {\xi ,Q} \right) = \sum\limits_r {\alpha _r
\rho _r \left( {\xi ,Q} \right)}\ .
\end{equation}
It is important, however, that our \emph{dynamically} derived
randomness and probability need not rely on such ``statistical",
empirically based result, so that the basic expressions
(\ref{eq:14})--(\ref{eq:18}) remain valid even for a \emph{single}
event of realisation emergence and \emph{before} any event happens
at all.
\index{a priori probability}

Realisation probability distribution can be obtained in another way,
involving \emph{generalised wavefunction} and \emph{Born's
probability rule}
\cite{Kir:USciCom,Kir:SymCom,Kir:RealQuMach,Kir:Conscious,Kir:QFM,Kir:75Wavefunc}.
The wavefunction describes system state during its transition
between ``regular", localised realisations and constitutes a
particular, ``intermediate" realisation with extended and ``loose"
(chaotically changing) structure, where system components
transiently disentangle before forming the next ``regular"
realisation. The intermediate, or ``main", realisation is
\emph{explicitly obtained} in the unreduced EP formalism as the
single, \emph{exceptional} one for which the nonintegrable terms of
the general EP (\ref{eq:11}),(\ref{eq:16}) become indeed small and
it is reduced to a separable version of perturbative, ``mean-field"
type (\ref{eq:13})
\cite{Kir:USciCom,Kir:SymCom,Kir:RealQuMach,Kir:Conscious,Kir:QFM,Kir:75Wavefunc}.
This special realisation provides, in particular, the \emph{causal,
realistic} version of the \emph{quantum-mechanical wavefunction} at
the \emph{lowest}, quantum levels of complexity. The ``Born
probability rule", now causally derived and extended to any level of
world dynamics, states that realisation probability $\alpha _r$ is
determined by wavefunction value (its squared modulus for
``wave-like" complexity levels) for the respective system
configuration $X_r$: $\alpha _r  = \left| {\Psi \left( {X_r }
\right)} \right|^2$. The generalised wavefunction (or distribution
function) $\Psi(x)$ satisfies the \emph{universal Schr\"odinger
equation} (Sect.~\ref{SymCom}), rigorously \emph{derived} by
\emph{causal quantisation} of complex dynamics, while Born's
probability rule follows from the above \emph{dynamic} ``matching
conditions" for the state-function (\ref{eq:15}), which are
satisfied during transitions from regular realisations to the
wavefunction and back
\cite{Kir:USciCom,Kir:SymCom,Kir:RealQuMach,Kir:Conscious,Kir:QFM,Kir:75Wavefunc}.
It is \emph{only} this ``averaged", weak-interaction state of the
wavefunction, or ``main" realisation, that remains in the
\emph{single-valued} model and paradigm of unitary science, which
explains both its partial success and basic limitations.

Closely related to dynamic redundance is \emph{dynamic entanglement}
of interacting components, described in (\ref{eq:15}) by the
weighted products of state-function elements depending on various
degrees of freedom ($\xi, Q$). It is a \emph{rigorous} expression of
the tangible \emph{quality} of the emerging system structure, absent
in unitary models. The obtained \emph{dynamically multivalued
entanglement} describes a ``living" structure, permanently changing
and probabilistically \emph{adapting} its configuration, which
endows ``bio-inspired" and ``autonomic" technologies with a
\emph{well-specified basis}. The properties of dynamically
multivalued entanglement and adaptability are amplified due to
\emph{probabilistic fractality} of the unreduced problem solution
\cite{Kir:USciCom,Kir:Fract:1,Kir:RealQuMach,Kir:Fract:2,Kir:Conscious},
essentially extending usual, single-valued fractality and obtained
by application of the same EP method to solution of the truncated
system of equations (\ref{eq:12}) used in the first-level EP
expression (\ref{eq:11}),(\ref{eq:16}).
\index{dynamic entanglement}

We can now consistently and \emph{universally} define the unreduced
\emph{dynamic complexity}, $C$, of \emph{any} real system (or
interaction process) as arbitrary growing function of the total
number, $N_\Re$, of \emph{explicitly obtained} system realisations
or the rate of their change, $C = C\left( {N_\Re} \right),\ \ {{dC}
\mathord{\left/ {\vphantom {{dC} {dN_\Re   > 0}}} \right.
\kern-\nulldelimiterspace} {dN_\Re   > 0}}$, equal to zero for the
\emph{unrealistic} case of only one system realisation, $C\left(
{\rm{1}} \right){\rm{ = 0}}$. Suitable examples are provided by
$C\left( {N_\Re  } \right) = C_0 \ln N_\Re$, generalised energy/mass
(temporal rate of realisation change), and momentum (spatial rate of
realisation emergence)
\cite{Kir:USciCom,Kir:SelfOrg,Kir:SymCom,Kir:Fract:1,Kir:RealQuMach,Kir:Fract:2,Kir:Nano,Kir:Conscious,Kir:CommNet,Kir:SustTrans,Kir:QFM,Kir:Cosmo,Kir:QuChaos}.
It becomes clear now that the \emph{whole} dynamically single-valued
paradigm and results of the canonical theory (including its versions
of ``complexity" and e.g. ``multi-stability") correspond to exactly
\emph{zero} value of the unreduced dynamic complexity, which is
equivalent to the effectively zero-dimensional, point-like
projection of reality from the ``exact-solution" perspective (cf.
\cite{Perplexity,EndScience,Ellis:1,Ellis:2}).
\index{dynamic complexity}

Correspondingly, \emph{any} dynamically single-valued ``model" is
\emph{strictly regular} and \emph{cannot} possess any true,
intrinsic randomness (chaoticity), which can only be introduced
artificially, e.g. as a \emph{regular} ``amplification" of a
``random" (by convention) \emph{external} ``noise". By contrast, our
unreduced dynamic complexity is practically synonymous to the
equally universal and intrinsic \emph{chaoticity}, since
\emph{multiple} system realisations appearing and disappearing in
the \emph{real} space (and thus \emph{forming} its tangible,
changing structure) are \emph{redundant} (mutually
\emph{incompatible}), which is the origin of \emph{both} complexity
and chaoticity. The genuine dynamical chaos thus obtained has a
complicated internal structure (contrary to ill-defined unitary
``stochasticity") and \emph{always} contains \emph{partial
regularity}, which is dynamically, inseparably entangled with truly
random elements.

The universal dynamic complexity, chaoticity, and related properties
involve the \emph{essential, or dynamic, nonlinearity} of the
unreduced problem solution and system behaviour. It is provided by
dynamical links of the developing interaction process, as they are
expressed in EP dependence on the eigen-solutions to be found (see
(\ref{eq:9})--(\ref{eq:11}),(\ref{eq:16})). It is the
\emph{dynamically emerging} and \emph{irreducible} nonlinearity,
since it appears inevitably even for a ``linear" initial problem
expression (\ref{eq:1})--(\ref{eq:2}),(\ref{eq:5}), whereas usual,
mechanistic ``nonlinearity" is but an \emph{imposed}, dispensable
\emph{imitation} of the essential EP nonlinearity. Essential
nonlinearity leads to the omnipresent \emph{dynamic instability} of
any system state (realisation), since both are determined by the
same dynamic feedback mechanism.

Universality of our description leads, in particular, to the unified
understanding of the whole diversity of dynamical regimes and
structures
\cite{Kir:USciCom,Kir:SelfOrg,Kir:RealQuMach,Kir:Nano,Kir:Conscious,Kir:CommNet,Kir:SustTrans}.
One standard, limiting case of complex (multivalued) dynamics,
called \emph{uniform, or global, chaos}, is characterised by
essentially different realisations with a homogeneous probability
distribution ($N_r \approx 1$, $\alpha _r  \approx {1
\mathord{\left/ {\vphantom {1 {N_\Re  }}} \right.
\kern-\nulldelimiterspace} {N_\Re }}$ for all $r$ in (\ref{eq:18}))
and occurs when major parameters of interacting entities (suitably
represented by frequencies) have close values (which leads to a
strong ``conflict of interests" and resulting ``deep disorder"). The
complementary limiting regime of \emph{multivalued
self-organisation, or self-organised criticality (SOC)} emerges for
sufficiently different parameters of interaction components, so that
a small number of relatively rigid, low-frequency components
``enslave" a hierarchy of high-frequency and rapidly changing, but
configurationally similar, realisations (i.e. $N_r \sim N_\Re$ and
realisation probability distribution is highly uneven). The
difference of this extended, multivalued self-organisation (and SOC)
from usual, unitary version is essential: despite the rigid
\emph{external} shape of system configuration in this regime, it
contains the intense ``internal life" and \emph{chaos} of
\emph{permanently} changing ``enslaved" realisations (which are
\emph{not} superposable unitary ``modes"). In this sense the
generalised SOC structure, and with it the whole unreduced
complexity, can be described as \emph{confined chaos}, where global
chaos has the lowest and quasi-regular SOC the highest degree of
chaos confinement.
\index{chaos}
\index{self-organisation}

Another advance with respect to unitary ``science of complexity" is
that the unreduced, multivalued self-organisation \emph{unifies} the
\emph{essentially extended} versions of a whole series of separated
unitary ``models", including ``self-organisation", ``synergetics",
SOC, any ``synchronisation", ``control of chaos", ``attractors", and
``mode locking". All intermediate dynamic regimes between those two
limiting cases of uniform chaos and quasi-regular SOC, as well as
their multi-level combinations, are obtained for respective
parameter values.

The point of transition to the strong chaos is expressed by the
\emph{universal criterion of global chaos onset}:
\begin{equation}\label{eq:20}
\kappa  \equiv \frac{{\Delta \eta _i }}{{\Delta \eta _n }} =
\frac{{\omega _\xi  }}{{\omega _q }} \cong 1\ ,
\end{equation}
where $\kappa$ is the introduced \emph{chaoticity} parameter,
$\Delta \eta _i$, $\omega _\xi$ and $\Delta \eta _n \sim \Delta
\varepsilon$, $\omega _q$ are energy-level separations and
frequencies for inter-component and intra-component motions,
respectively. At $\kappa \ll 1$ one has an externally regular
multivalued SOC regime, which degenerates into global chaos as
$\kappa$ grows from 0 to 1, and maximum irregularity at $\kappa
\approx 1$ is again transformed into a SOC kind of structure (but
with a ``reversed" configuration) at $\kappa \gg 1$.

One can compare this transparent and universal picture with the
existing diversity of separated and incomplete unitary criteria of
chaos and regularity. Only the former provide a real possibility of
understanding and control of ICT systems of arbitrary complexity,
where more regular, SOC regimes can serve for (loose) control of
system dynamics, while less regular ones can also play a
\emph{positive} role of efficient search and adaptation means. This
combination forms the basis of any ``biological" and ``intelligent"
kind of behaviour
\cite{Kir:USciCom,Kir:Fract:1,Kir:RealQuMach,Kir:Fract:2,Kir:Conscious,Kir:CommNet,Kir:SustTrans}
and therefore can constitute the essence of \emph{intelligent ICT
paradigm} supposed to extend the now realised (quasi-) regular kind
of operation in the uttermost limit of SOC ($\kappa \to 0$). While
the latter \emph{inevitably} becomes inefficient with growing system
sophistication (where the chaos-bringing resonances of (\ref{eq:20})
\emph{cannot} be avoided), it definitely lacks the ``intelligent
power" of unreduced complex dynamics to generate meaning and
adaptable structure development.

\subsection{Huge Efficiency of Unreduced Complex Dynamics and Universal
Symmetry of Complexity}\label{SymCom}
\emph{Dynamically probabilistic fractality} is the intrinsic
property of unreduced interaction development
\cite{Kir:USciCom,Kir:Fract:1,Kir:RealQuMach,Kir:Fract:2,Kir:Conscious}.
It is obtained by application of the same EP method
(\ref{eq:9})--(\ref{eq:11}) to the truncated system of equations
(\ref{eq:12}), then to the next truncated system, etc., which gives
the irregular and \emph{probabilistically adapting} hierarchy of
realisations showing the intermittent mixture of global chaos and
regularity, or \emph{confined randomness} (Sect.~\ref{UniCom}). The
total realisation number $N_\Re$, and thus \emph{operation power},
of this autonomously branching interaction process with a
\emph{dynamically parallel} structure grows \emph{exponentially}
within any time period. It can be estimated in the following way
\cite{Kir:RealQuMach,Kir:Fract:2,Kir:Nano,Kir:Conscious,Kir:CommNet}.
\index{probabilistic fractality}

If our system of inter-connected elements contains $N_{{\rm{unit}}}$
``processing units", or ``junctions", and if each of them has
$n_{{\rm{conn}}}$ real or ``virtual" (possible) links, then the
total number of interaction links is $N = n_{{\rm{conn}}}
N_{{\rm{unit}}}$. In most important cases $N$ is a huge number: for
both human brain and genome interactions $N$ is greater than
$10^{12}$, and being much more variable for communication/software
systems, it can easily grow to similar ``astronomical" ranges. The
key property of \emph{unreduced, complex} interaction dynamics,
distinguishing it from any unitary version, is that the maximum
number $N_\Re$ of realisations taken by the system (also per time
unit) and determining its real ``power" $P_{{\rm{real}}}$ (of
search, memory, cognition, etc.) is given by the number of \emph{all
possible combinations of links}, i.e.
\begin{equation}\label{eq:21}
P_{{\rm{real}}}  \propto N_\Re = N! \to \sqrt {2{\rm{\pi }}N} \left(
{\frac{N}{e}} \right)^N  \sim N^N  \gg  \gg N\ .
\end{equation}
Any unitary, sequential model of the same system (including its
\emph{mechanistically} ``parallel" and ``complex" modes) would give
$P_{{\rm{reg}}} \sim N^\beta$, with $\beta \sim 1$, so that
\begin{equation}\label{eq:22}
P_{{\rm{real}}}  \sim \left( {P_{{\rm{reg}}} } \right)^N  \gg  \gg
P_{{\rm{reg}}}  \sim N^\beta\ .
\end{equation}
Thus, for $N \sim 10^{12}$ we have $P_{{\rm{real}}} \gg 10^{10^{13}
} \gg 10^{10^{12} } \sim 10^N  \to \infty $, which is a ``practical
infinity", also with respect to the unitary power of $N^\beta \sim
10^{12}$.

These estimates demonstrate the true power of complex (multivalued)
communication dynamics that remains suppressed within the now
dominating unitary, quasi-regular operation mode. Huge power of
complex-dynamical interaction correlate with the new \emph{quality}
emergence, such as \emph{intelligence} and \emph{consciousness} (at
higher levels of complexity) \cite{Kir:RealQuMach,Kir:Conscious}, in
direct relation to our \emph{intelligent} communication paradigm
meaning that such properties as \emph{sensible}, context-related
information processing, personalised \emph{understanding} and
autonomous \emph{creativity} (useful self-development), desired for
the new ICT systems, are \emph{inevitable} qualitative
manifestations of the above ``infinite" power.

Everything comes at a price, however, and a price to pay for the
above qualitative advantages is rigorously specified as
\emph{irreducible dynamic randomness} and thus unpredictability of
operation details of complex information-processing systems. We
rigorously confirm here an evident idea that \emph{autonomous}
adaptability and genuine \emph{creativity} exclude any detailed,
regular programming in principle. But then what can serve as a
guiding principle and practical construction strategy for those
qualitatively new communications networks and their intelligent
elements? We show that guiding rules and strategy are determined by
a general law of real dynamics, the \emph{universal symmetry, or
conservation, of complexity}
\cite{Kir:USciCom,Kir:SymCom,Kir:Fract:1,Kir:RealQuMach,Kir:Fract:2,Kir:Conscious,Kir:CommNet,Kir:SustTrans,Kir:QFM,Kir:Cosmo}.
This universal ``order of nature" unifies the extended versions of
all usual (correct) laws, symmetries, and principles (now
\emph{causally derived} and \emph{realistically} interpreted).
Contrary to any unitary symmetry, the universal symmetry of
complexity is \emph{irregular} in its structure, but always
\emph{exact} (never ``broken"). Its ``horizontal" manifestation (at
a given complexity level) implies \emph{dynamic transformation} of
the system between its changing realisations, as opposed to
\emph{abstract} ``symmetry operator" idea. Therefore the symmetry of
system complexity totally determines its dynamics and expresses the
deep connection between often visibly dissimilar and chaotically
changing configurations.

Another, ``vertical" manifestation of the symmetry of complexity
determines emergence and development of \emph{different} complexity
levels of a real interaction. System ``potentiality", or \emph{real}
power to create \emph{new} structure is universally described by a
form of complexity called \emph{dynamic information} and
generalising usual ``potential energy"
\cite{Kir:USciCom,Kir:SymCom,Kir:RealQuMach,Kir:Conscious,Kir:Cosmo}.
This potential, latent complexity is transformed, during interaction
development, into explicit, ``unfolded" form of \emph{dynamic
entropy} (generalising kinetic, or heat, energy). Universal
\emph{conservation of complexity} means that this important
transformation, determining every system dynamics and evolution,
\emph{preserves} the \emph{sum} of dynamic information and entropy,
or \emph{total complexity} (for a given system or process). This
universal formulation of the symmetry of complexity includes its
above ``horizontal" manifestation and, for example, extended and
\emph{unified} versions of the first and second laws of
thermodynamics (i.e. conservation of energy \emph{by} its
\emph{permanent} degradation). It also helps to eliminate persisting
series of confusions around information, entropy, and complexity in
the unitary theory (thus, any real, useful ``information" is
expressed rather by our dynamic entropy
\cite{Kir:USciCom,Kir:RealQuMach}).\index{symmetry of complexity}

It is not difficult to show
\cite{Kir:USciCom,Kir:SymCom,Kir:RealQuMach,Kir:Conscious,Kir:QFM,Kir:Cosmo}
that a universal measure of dynamic information is provided by
action $\cal A$ known from classical mechanics, but now acquiring a
universal, essentially nonlinear and causally complete meaning. One
obtains then the universal expression of complexity conservation law
in the form of generalised Hamilton-Jacobi equation for ${\cal A} =
{\cal A} (x,t)$:
\begin{equation}\label{eq:23}
\frac{{\Delta \cal A}}{{\Delta t}}\left| {_{x = {\rm const}} }
\right. + H\left( {x,\frac{{\Delta \cal A}}{{\Delta x}}\left| {_{t =
{\rm const}} ,t} \right.} \right) = 0\ ,
\end{equation}
where the \emph{Hamiltonian}, $H = H(x,p,t)$, considered as a
function of \emph{emerging} space coordinate $x$, momentum $p =
\left( {{{\Delta \cal A} \mathord{\left/ {\vphantom {{\Delta A}
{\Delta x}}} \right. \kern-\nulldelimiterspace} {\Delta x}}}
\right)\left| {_{t = {\rm const}} } \right.$, and time $t$,
expresses the unfolded, entropy-like form of differential
complexity, $H = \left( {{{\Delta S} \mathord{\left/ {\vphantom
{{\Delta S} {\Delta t}}} \right. \kern-\nulldelimiterspace} {\Delta
t}}} \right)\left| {_{x = {\rm const}} } \right.$ (note that
discrete, rather than continuous, versions of derivatives here
reflect the \emph{quantised} character of unreduced complex dynamics
\cite{Kir:USciCom,Kir:SymCom,Kir:RealQuMach,Kir:Conscious,Kir:QFM,Kir:Cosmo}).
As in the naturally \emph{dualistic} multivalued dynamics every
localised, ``regular" realisation is transformed into the extended
wavefunction and back (Sect.~\ref{UniCom}), one obtains also the
universal Schr\"odinger equation for the generalised wavefunction
(or distribution function) ${\mit \Psi} (x,t)$ by applying causal
quantisation procedure
\cite{Kir:USciCom,Kir:SymCom,Kir:RealQuMach,Kir:Conscious,Kir:QFM,Kir:Cosmo}
to the Hamilton-Jacobi equation (\ref{eq:23}):
\begin{equation}\label{eq:24}
{\cal A}_0 \frac{{\partial {\mit \Psi} }}{{\partial t}} = \hat
H\left( {x,\frac{\partial }{{\partial x}},t} \right){\mit \Psi}\ ,
\end{equation}
where ${\cal A}_0$ is a characteristic action value (equal to
Planck's constant at the lowest, quantum levels of complexity) and
the Hamiltonian operator, $\hat H$, is obtained from the Hamiltonian
function $H = H(x,p,t)$ of equation (\ref{eq:23}) with the help of
causal quantisation (we put continuous derivatives here for
simplicity).
\index{causal quantisation}

Equations (\ref{eq:23})--(\ref{eq:24}) represent the universal
differential expression of the symmetry of complexity showing how it
determines dynamics and evolution of any system or interaction
process (they also justify our use of the Hamiltonian form for the
starting existence equation (\ref{eq:1})--(\ref{eq:2})). This
universally applicable Hamilton-Schr\"odinger formalism can be
useful for rigorous description of any complex interaction network,
provided we find its \emph{truly complete} (dynamically multivalued)
solution with the help of unreduced EP method (Sect.~\ref{UniCom}).
\index{Hamilton-Schr\"odinger formalism}

\section{Unified Science of Complex ICT Systems}\label{SciComNet}
\subsection{Main Principles of Complex ICT System Operation and Design}\label{Principles}
The rigorously derived framework of the Universal Science of
Complexity (Sect.~\ref{ComDyn}) finds its further confirmation in
numerous applications at different complexity levels, from
fundamental physics and cosmology (Quantum Field Mechanics)
\cite{Kir:USciCom,Kir:RealQuMach,Kir:QFM,Kir:Cosmo,Kir:QuChaos,Kir:75Wavefunc}
to living organism dynamics (causally specified genomics and
nanobiotechnology)
\cite{Kir:USciCom,Kir:Fract:1,Kir:RealQuMach,Kir:Fract:2,Kir:Nano},
ecological system development (realistic sustainability concept)
\cite{Kir:USciCom,Kir:SustTrans}, and theory of emergent true
intelligence and consciousness
\cite{Kir:USciCom,Kir:RealQuMach,Kir:Conscious} (see also
Sect.~\ref{NewMath}). These results give a realistic hope for an
equally successful application of the same complexity concept to the
new generation of communication and software systems with
``bio-inspired" and ``intelligent" properties
\cite{WAC,FP6,Bull,DiMarzo,SemWeb,ManyOne}, which are actually
\emph{indispensable} for efficient work with the \emph{critically
emerging} real-system complexity from the above applications
(Sect.~\ref{Intro}). We provide here an outline of the \emph{main
principles} of expected behaviour and design of complex-dynamic
communication networks and software as they follow from the
universal complexity framework (Sect.~\ref{ComDyn}).
\index{complex-dynamic communication}

The \emph{single unifying principle} of complex system dynamics and
evolution is provided by the \emph{universal symmetry of complexity}
describing complexity \emph{conservation} by its \emph{permanent
transformation} from dynamic information into entropy as the
\emph{unified structure and purpose} of \emph{any system evolution}
(Sect.~\ref{SymCom}). While the very existence of such unified law
is important for efficient analysis of generalised autonomic
networks (Sect.~\ref{Intro}), we can specify now the ensuing
particular principles that can be especially useful for their
practical design and control.

We start with the \emph{complexity correspondence principle} that
directly follows from the universal symmetry of complexity and takes
various forms for different application tasks
\cite{Kir:USciCom,Kir:RealQuMach,Kir:Conscious}. A general enough
formulation maintains that any interaction between complex systems
(e.g. within a ``global" system) tend to have \emph{maximum
efficiency} for \emph{comparable} values of interacting system
complexities (see Sect.~\ref{UniCom} for the universal complexity
definition). Moreover, interaction components with higher complexity
tend to ``enslave", or control, those with lower complexity within a
resulting SOC-type state (Sect.~\ref{UniCom}), while very close
complexities of interacting components often give rise to global
chaoticity.
\index{complexity correspondence principle}

It follows that in order to increase efficiency one should use tools
of certain complexity for control of structures of \emph{comparable
and slightly lower} complexity. Lower-complexity tools \emph{cannot}
correctly control or \emph{even simulate} higher-complexity
behaviour in principle, while a \emph{much} higher complexity tool
will produce a lot of \emph{unnecessary} activity during control of
low-complexity structure. This ``simple" rule has a \emph{rigorous,
reliable} basis and helps to avoid any inefficient solution (e.g.
usual \emph{unconditional}, ``total" elimination of randomness).

Reproduction, or simulation, of a system behaviour by another,
``computing" or controlling system can also be successful
\emph{only} if the simulating system has \emph{superior} complexity,
which immediately shows, for example, that any quantum device could
at best simulate or compute only another quantum, but \emph{not}
classical, localised and deterministic behaviour, which provides a
\emph{rigorous} proof of \emph{impossibility} of real quantum
computation \cite{Kir:USciCom,Kir:RealQuMach}. Losses produced by
the dominating neglect of underlying hierarchy of complexity are
evident and include a similar situation in nanotechnology (cf.
\cite{Kir:Nano}). Another use of complexity correspondence involves
popular ideas of ``context-based" information technologies, where
the necessity of \emph{unreduced dynamic complexity} in ICT systems
becomes evident, as any \emph{human} ``context" has a high enough
complexity.

The above rule of control (or ``enslavement") of lower-complexity
dynamics by a higher-complexity tool can be extended to a general
\emph{principle of complex-dynamical control}. Contrary to unitary
control schemes (e.g. ``control of chaos"), the \emph{realistically}
substantiated complex-dynamical control paradigm shows that any
resulting, ``controlled" dynamics \emph{cannot} be regular, i.e.
``totally controlled", as it is implied by the unitary control idea
leading to unpredictably (and \emph{inevitably}) emerging
\emph{catastrophic} failures of any technological systems with
usual, ``protective" control design. In reality one \emph{always}
obtains a \emph{dynamically multivalued}, internally \emph{chaotic}
SOC state, and the \emph{general purpose} of complex-dynamic,
reality-based control is to ensure optimal, quasi-\emph{free}
\emph{development of the global system complexity}, including
gentle, ``orienting" actions of control that cannot be separated
from the controlled system dynamics and should be considered within
a \emph{unified, unreduced interaction analysis}
(Sect.~\ref{ComDyn}). In that way one can realistically obtain a
\emph{failure-proof, catastrophe-free systems} that will avoid
\emph{big} crashes by using \emph{creative} power of \emph{small},
interaction-driven irregularities, quite similar to \emph{unreduced
life dynamics}, now \emph{causally} understood
\cite{Kir:USciCom,Kir:Fract:1,Kir:RealQuMach,Kir:Fract:2}.
\index{complex-dynamical control}

This brings us to the principle of \emph{huge creative power of
unreduced complex dynamics} as it is described above
(Sect.~\ref{SymCom}). It implies, practically, that designing truly
autonomous and intelligent ICT systems, one should ``liberate" them
to go \emph{freely}, by their own way, unpredictable in its
\emph{chaotically varying details}, to the \emph{well-defined
general purpose} of \emph{maximum complexity-entropy} obtained at
the expense of \emph{inserted dynamical information} that replaces
usual deterministic programme. One obtains thus the exponentially
huge, practically infinite gain in efficiency with respect to
unitary, sequential operation due to interactive adaptability of
\emph{probabilistic fractal} of \emph{dynamically emerging} links
\cite{Kir:RealQuMach,Kir:Fract:2,Kir:Nano,Kir:Conscious,Kir:CommNet}
(Sect.~\ref{SymCom}). In exchange, one should accept the
omnipresent, massive, unavoidable \emph{uncertainty} of unreduced
interaction dynamics. It can, however, be properly \emph{confined}
and \emph{constructively} used by alternating uniformly chaotic
(irregular, ``searching") and SOC (regular, ``fixing") states, whose
separation is governed by our \emph{unified chaoticity criterion}
(\ref{eq:20}) in terms of major system resonances.
\index{adaptability}
\index{fractal}

In conclusion of this outline of major principles of complex
autonomic networks, it would be useful to return to the global
framework of the symmetry of complexity that shows, in the above
way, \emph{what can happen} in the strongly interactive ICT system
of \emph{arbitrary complexity} and how one can efficiently design
and control the unreduced interaction results by its \emph{causally
complete understanding}. The latter can certainly be properly
specified and adapted, where necessary, to any particular case of
reduced, mechanistic ``complexity", but now with the underlying
clear understanding of the performed actions
(Sect.~\ref{Conclusion}).

\subsection{Complexity Transition in ICT Systems: Towards the New Era of Intelligent
Communication Technology}\label{ComTrans}
The \emph{qualitatively big} transition from unrealistic unitary
models to the unreduced, multivalued dynamics of real, massively
interactive ICT systems can be designated as \emph{complexity
transition}. It has a narrow meaning of transition from the
uttermost limit of pseudo-regular SOC of usual ICT to the fractal
dynamic hierarchy of various chaotic states, according to the
chaoticity criterion (\ref{eq:20}). In a wider sense, one deals with
a fundamentally based \emph{change of concept} of modern technology
(Sects.~\ref{Intro},~\ref{Principles},~\ref{Conclusion}) and related
\emph{way of development}.
\index{complexity transition}

Unreduced complexity appearance should rather be tested first at the
level of \emph{software}. Using the results of complexity
correspondence principle described in the previous section, we can
suppose that complexity transition can be conveniently started
within \emph{context-based} technology, where complexity-bringing
interaction involves essential, \emph{structure-changing}, autonomic
exchange between context-bearing elements. Their unreduced
interaction should then be designed according to the principles of
complex interaction dynamics
(Sects.~\ref{ComDyn},~\ref{Principles}).

As the unreduced interaction complexity forms a growing hierarchy of
levels \cite{Kir:USciCom,Kir:RealQuMach,Kir:Conscious}, one obtains
eventually a whole series of system transitions to ever growing
complexity. Software version of initial complexity emergence will
later involve \emph{hardware} elements into structure-changing
interaction processes. Communication network or its respective parts
operate in that case as a single, \emph{holistic} process of
``generalised quantum beat" (chaotic realisation change). Transition
to a \emph{high enough} complexity level will bring about first
elements of genuine network \emph{intelligence} and then
\emph{consciousness}, as both these properties can be consistently
explained as high enough levels of unreduced interaction complexity
\cite{Kir:USciCom,Kir:RealQuMach,Kir:Conscious}. One can designate
then \emph{intelligence and consciousness transitions} as
sufficiently high-level complexity transitions involving unreduced
soft- and hardware interaction. Whereas lower-level complexity
transitions can be limited to separate network parts and operation
layers, such higher-level features as intelligence and (machine)
consciousness will progressively involve the \emph{whole} network
dynamics, which is the evident \emph{highest} level of communication
network \emph{autonomy}. Whereas already the lowest complexity
transition involves context-bearing elements of human complexity,
intelligence transition marks the beginning of \emph{inseparable
entanglement} of machine \emph{and} human complexity development
that can be the \emph{unique} real way of progressive development of
``natural" intelligence and consciousness
\cite{Kir:Conscious,Kir:CommNet}.

\section{New Mathematics of Complexity and Its Applications}\label{NewMath}
After having outlined, in the previous section, practically oriented
principles of ICT applications of the unreduced complexity, let us
now summarise purely mathematical, rigorously expressed novelties of
the unreduced interaction analysis to be used in the new kind of
knowledge (see also the end of paper \cite{Kir:Fract:2}).

The \emph{new mathematics of complexity} is represented by the
\emph{unified, single structure} of \emph{dynamically probabilistic
fractal} obtained as the \emph{unreduced solution} of \emph{real}
interaction problem (Sect.~\ref{UniCom}). All its properties,
describing the \emph{exact} world structure and dynamics, are
unified within the single, never broken \emph{symmetry, or
conservation, of complexity} including its \emph{unceasing
transformation} from complexity-information to complexity-entropy
(Sect.~\ref{SymCom}).
\index{mathematics of complexity}

One can emphasize several features of this unified structure and law
of the new mathematics, distinguishing it essentially from unitary
mathematics:

\begin{list}{(\roman{enumi})}{\usecounter{enumi}}

\item
\emph{Non}uniqueness of any real (interaction) problem solution
taking the form of its \emph{dynamic multivaluedness (redundance)}.
Exclusively \emph{complex-dynamic} (multivalued, internally
\emph{chaotic}) existence of any real system (cf. conventional
``existence and uniqueness" theorems).

\item
Omnipresent, explicit \emph{emergence} of \emph{qualitatively new}
structure and \emph{dynamic origin of time} (change) and
\emph{events}: $a \ne a$ for \emph{any} structure/element $a$ in the
new mathematics \emph{and} reality, while $a = a$ (self-identity
postulate) in the \emph{whole} usual mathematics, which thus
\emph{excludes any real change} in principle.\index{origin of time}

\item
Fractally structured \emph{dynamic entanglement} of unreduced
problem solution (interaction-driven structure weaving within any
single realisation): \emph{rigorous} expression of \emph{material
quality} in mathematics (as opposed to ``immaterial", qualitatively
``neutral", ``dead" structures of usual mathematics).

\item
Basic \emph{irrelevance} of perturbation theory and ``exact
solution" paradigm: the unreduced problem solution is
\emph{dynamically random} (permanently, chaotically changing),
\emph{dynamically entangled} (internally textured and ``living") and
\emph{fractal} (hierarchically structured). \emph{Unified dynamic
origin} and \emph{causally specified} meaning of
\emph{nonintegrability}, \emph{nonseparability},
\emph{noncomputability}, \emph{randomness}, \emph{uncertainty
(indeterminacy)}, \emph{undecidability}, \emph{``broken symmetry"},
etc. Real interaction problem is nonintegrable and nonseparable
\emph{but} solvable. Realistic mathematics of complexity is
\emph{well defined} (\emph{certain}, \emph{unified} and
\emph{complete}), but its structures are intrinsically ``fuzzy"
(dynamically \emph{indeterminate}) and properly \emph{diverse}
(\emph{not} reduced to numbers or geometry).\index{noncomputability}

\item
\emph{Dynamic discreteness (causal quantisation)} of the unreduced
interaction products (realisations): \emph{qualitative}
inhomogeneity, or \emph{nonunitarity}, of any system structure and
evolution, \emph{dynamic} origin of (fractally structured)
\emph{space}. Qualitative \emph{irrelevance} of usual unitarity,
continuity \emph{and} discontinuity, calculus, and \emph{all} major
structures (evolution operators, symmetry operators, \emph{any}
unitary operators, Lyapunov exponents, path integrals,
etc.).\index{dynamic discreteness}

\end{list}

Let us recall now how these \emph{fundamental} novelties of the
universal science of complexity help to solve consistently
\emph{real-world problems}
\cite{Kir:USciCom,Kir:SelfOrg,Kir:SymCom,Kir:Fract:1,Kir:RealQuMach,Kir:Fract:2,Kir:Nano,Kir:Conscious,Kir:CommNet,Kir:SustTrans,Kir:QFM,Kir:Cosmo,Kir:QuChaos,Kir:Channel,Kir:75Wavefunc}
that accumulate and remain \emph{unsolvable} within the unitary
science paradigm:

\begin{list}{(\arabic{enumi})}{\usecounter{enumi}}

\item
In \emph{particle and quantum physics} one obtains \emph{causal},
\emph{unified} origin and structure of \emph{elementary particles},
\emph{all} their \emph{properties} (``intrinsic", quantum,
relativistic), and fundamental \emph{interactions}
\cite{Kir:USciCom,Kir:RealQuMach,Kir:QFM,Kir:Cosmo,Kir:75Wavefunc}.
\emph{Complex-dynamic origin of mass} avoids any additional,
abstract entities (Higgs bosons, zero-point field, extra dimensions,
etc.). \emph{Renormalised Planckian units} provide consistent
\emph{mass spectrum} and other problem solution.
\emph{Complex-dynamic cosmology} resolves the dark mass and energy
problems without ``invisible" entities, together with other old and
new problems of unitary cosmology.

\item
At a higher complexity sublevel of \emph{interacting particles}
\cite{Kir:USciCom,Kir:RealQuMach,Kir:QFM,Kir:QuChaos,Kir:Channel}
one obtains \emph{genuine, purely dynamic quantum chaos} for
Hamiltonian (nondissipative) dynamics and \emph{correct
correspondence principle}. A slightly dissipative interaction
dynamics leads to the \emph{causally complete} understanding of
\emph{quantum measurement} in terms of \emph{quantum} dynamics.
\emph{Intrinsic classically} emerges as a \emph{higher complexity
level} in a \emph{closed}, bound system, like atom.

\item
\emph{Realistic, causally complete} foundation of
\emph{nanobiotechnology} is provided by \emph{rigorous} description
of \emph{arbitrary} nanoscale interaction, revealing the
\emph{irreducible} role of \emph{chaoticity}
\cite{Kir:RealQuMach,Kir:Nano}. \emph{Exponentially huge power} of
unreduced, complex nanobiosystem dynamics explains the
\emph{essential properties of life} and has direct relation to
complex ICT system development (Sect.~\ref{SymCom}).

\item
\emph{Causally complete} description of \emph{unreduced genomic
interactions} leads to \emph{reliable, rigorously substantiated
genetics} and consistent understanding of related \emph{evolutionary
processes} \cite{Kir:Fract:1,Kir:Fract:2}.

\item
Higher-complexity applications include \emph{general many-body
problem solution} and related description of ``difficult" cases in
\emph{solid-state physics}, unreduced dynamics and evolution of
\emph{living organisms}, \emph{integral medicine}, emergent
(genuine) \emph{intelligence} and \emph{consciousness},
\emph{complex ICT system dynamics}, \emph{creative ecology} and
practically efficient \emph{sustainable development concept},
\emph{rigorously} specified \emph{ethics} and \emph{aesthetics}
\cite{Kir:USciCom,Kir:Fract:1,Kir:RealQuMach,Kir:Fract:2,Kir:Conscious,Kir:CommNet,Kir:SustTrans}.

\end{list}

Note that only the unreduced, universal concept of complexity can be
useful for real problem solution culminating in creation of complex,
autonomic and intelligent ICT systems, which in their turn form a
\emph{consistent}, necessary basis for further control and
development of real interaction complexity.

\section{Conclusion and Perspectives}\label{Conclusion}
We have presented a rigorously specified, working prototype of the
unified science of complex ICT systems demonstrating its necessity,
feasibility and practical application efficiency. One obtains thus a
new, intrinsically unified and realistic kind of knowledge with
extended possibilities of consistent understanding and progressive
development of real-world complexity.

Application of the universal science of complexity to autonomic
communication and information systems has a \emph{special
importance} among other applications, since it is the \emph{first}
case of totally \emph{artificial}, man-made systems that can possess
\emph{unreduced complexity} features comparable to those of natural
systems and remaining irrational ``mysteries" within the unitary
science framework (starting from ``quantum mysteries"
\cite{Kir:USciCom,Kir:RealQuMach,Kir:QFM,Kir:Cosmo,Kir:QuChaos,Kir:Channel,Kir:75Wavefunc}).
Successful realisation of unreduced ICT complexity will open the way
to a much larger, unlimited and now \emph{reliable} complexity
design becoming so necessary today because of the rapid
\emph{empirical} technology progress (Sect.~\ref{Intro}). Those
major purposes \emph{cannot} be attained within unitary
\emph{imitations} of communication and software complexity (e.g.
\cite{Bull,KocarevVattay}), since they avoid the \emph{unreduced},
network-wide interaction analysis and use \emph{essentially
simplified} models even for separate component description (in
particular, they cannot describe the emerging genuine chaoticity
\cite{Kir:USciCom,Kir:RealQuMach}).

It is difficult to have serious doubts about basic consistency of a
unified complexity framework based on the unreduced problem solution
and confirmed by a variety of applications
(Sects.~\ref{ComDyn},~\ref{NewMath}). Understanding of unreduced
interaction complexity is indispensable for efficient design of even
regular, but mechanically ``complicated" systems. It is evident that
all increasingly popular ``bio-inspired", autonomic and
``intelligent" \emph{imitations} of natural complexity will be much
more successful with the help of \emph{consistent understanding} of
the \emph{unreduced} versions and properties of life, intelligence,
etc. In this sense one can say that application of the universal
science of complexity will certainly provide \emph{consistent}
clarification of what is possible \emph{or} impossible in
\emph{artificial complexity design}, whereupon further development
of ICT applications of unreduced complexity analysis can produce
\emph{only positive} (and urgently needed) result.

%
%



\printindex

\end{document}